\documentstyle[aps,prl]{revtex}
\def\ltap{\ \raise.3ex\hbox{$<$\kern-.75em\lower1ex\hbox{$\sim$}}\ }
\def\gtap{\ \raise.3ex\hbox{$>$\kern-.75em\lower1ex\hbox{$\sim$}}\ }

\def\CL{{\cal L}}

\def\be{\begin{equation}}
\def\ee{\end{equation}}

\begin{document}
\twocolumn[
\widetext
\hfill\hbox{\vbox{\hbox{UW/PT-97/07}\hbox{hep-ph/9703379}}}
\vskip 10pt
{\large\bf\centering
Contact Terms, Compositeness, and Atomic Parity
Violation \\
}\vskip5pt

{\rm\centering  Ann E. Nelson \\}
\vskip1.5pt

{\small\it\centering\ignorespaces
           Department of Physics 1560, University of Washington,
           Seattle, WA 98195-1560, USA\\
           {\tt anelson@phys.washington.edu}\\
}

\begin{abstract} 
The most severe constraints on quark-lepton
four-fermion contact interactions come from the agreement of atomic
parity violation measurements with the Standard Model. In this letter
I note that for contact interactions which arise in theories of
composite quarks and leptons, other approximate global symmetries than
parity
can eliminate the contribution  of contact terms to atomic parity
violation. 
The most stringent tests of compositeness therefore come from the 
high energy collider experiments at LEP II, HERA, and the Tevatron. 
\end{abstract}
\vskip.5pc

]
\narrowtext

Recently, there has been great interest in four-fermion contact
interactions between quarks and leptons, as such terms might account
for the reported excess of high $Q^2$ events in the HERA experiments
\cite{hone,zeus,alt,babu,barger,kalinowski,GGN}. The  8 relevant terms are usually
written in the form
\be\label{contact}
\Delta \CL= \sum_{i,j=L,R;q=u,d}{4 \pi\eta^q_{ij}\over(\Lambda_{ij}^q)^2}
\bar e_i\gamma_\mu e_i \bar q_j\gamma^\mu q_j
\ee
 where $\eta^q_{ij}=\pm1$.

It is just possible to find such terms which can  account for the HERA excess
while still satisfying the constraints deduced from studies of 
$e^+ e^-\rightarrow hadrons$ \cite{opal} and $p\bar
p\rightarrow
e^+e^-X$ \cite{tevatron},  for $\Lambda\sim 3$ TeV
\cite{alt,babu,barger}.

Stronger limits on such contact terms arise from atomic parity
violation (APV) measurements \cite{langacker,leurer,
davidson}. A contact interaction apparently shifts the nuclear
weak charge $Q_W$ by an amount
\be \Delta Q_W=-2 [\Delta C_{1u}(2Z+N)+\Delta C_{1d}(2N+Z)]\ee
where
\be\label{weakcharge}\Delta C_{1q}={\sqrt{2}\pi\over G_F}\left(
{\eta^q_{RL}\over (\Lambda^q_{RL})^2
}-{\eta^q_{LR}\over (\Lambda^q_{LR})^2
}+{\eta^q_{RR}\over (\Lambda^q_{RR})^2
}-{\eta^q_{LL}\over (\Lambda^q_{LL})^2
}\right). \ee
 
If no cancellations in eq.~\ref{weakcharge}  occur amongst the various terms,
measurements of the weak charge of Cesium \cite{cesium} imply
$\Lambda$'s$\gtap 10$~TeV. Thus the bounds from atomic parity violation  on
quark-lepton contact terms appear to be much stronger than those from any
collider experiments. Several authors \cite{babu,barger}
  have invoked a new
parity conserving contact interaction in order to explain the HERA
data while avoiding  the APV constraint.
They therefore assume that
\begin{eqnarray}\label{parity}
{\eta^q_{RL}\over (\Lambda^q_{RL})^2
}&=&{\eta^q_{LR}\over (\Lambda^q_{LR})^2}\\
{\eta^q_{RR}\over (\Lambda^q_{RR})^2
}&=&{\eta^q_{LL}\over (\Lambda^q_{LL})^2}\ .
\end{eqnarray}
The theoretical
motivation for imposing the restrictions of eq.~\ref{parity} is
unclear. An awkward feature of eq.~\ref{parity} is that 
SU(2) gauge symmetry
makes it necessary to introduce a right handed neutrino in order  to
have parity invariant and  gauge invariant contact 
terms involving leptons.

One interesting class of models which will lead to contact terms at
low energies are theories of composite quarks and leptons. In such
theories there are new strong confining dynamics at a scale $\Lambda$.
Unbroken chiral global symmetries of the
strong dynamics explain why the  quark and lepton bound states are
much lighter than $\Lambda$ \cite{thooft}. Any contact terms produced
by the strong dynamics will respect its global symmetries. 
These chiral symmetries may
be explicitely broken by small effects, {\it e.g.} by weak gauge
interactions, however  small symmetry breaking terms do not affect the
conclusions of this note.

It is an easy matter to find plausible
approximate global symmetries, other than parity,  
which will ensure cancellations in eq.~\ref{weakcharge}\cite {chivran}.
For instance  consider an approximate global SU(12)
acting on all left handed first generation quark states.
 The left chiral fields
\be (u_L, d_L, u^c_L, d^c_L)\ee transform as a 12-plet $\psi_L$.

Assuming that the new strong dynamics respects 
such a symmetry, it could generate only an SU(12) singlet
combination of the 
operators in eq.~\ref{contact},
which can be written in the form
\begin{eqnarray}&& \sum_{i=L,R}{4 \pi\eta_{i}\over(\Lambda_{i})^2}\bar e_i\gamma_\mu
e_i
\bar\psi_L\gamma^\mu\psi_L
\\&&=\sum_{i=L,R;q=u,d}{4 \pi\eta_{i}\over(\Lambda_{i})^2}\bar e_i\gamma_\mu
e_i(\bar q_L\gamma^\mu q_L-\bar q_R\gamma^\mu q_R)\ .
\end{eqnarray} 
Thus the SU(12) symmetry guarantees that 
\be \label{cancel} 
{\eta^q_{iL}\over (\Lambda^q_{iL})^2
}=-{\eta^q_{iR}\over (\Lambda^q_{iR})^2}\ ,
\ee
and so there is a cancellation in the contribution to $Q_W$. 

The SU(12) symmetry still allows for a non zero contribution to 
 the parity violating weak coefficient $C_{2q}$ \cite{pdg}, 
however the experimental constraints on
this term are less severe. In any case, an SU(3) symmetry acting on
all the left handed first generation leptons \be
(\nu^e_L, e_L, e^c_L)\ee would eliminate this contribution as well.

Much stronger constraints on contact terms can be obtained by
considering flavor changing neutral current decays and muon number
violation.
However such constraints can be satisfied by contact
terms which respect a horizontal flavor symmetry, such as an
$SU(2)\times SU(2)$, where one SU(2)
acts on the first two quark generations and the other on the first two
lepton generations. 

In summary, I have shown that composite models of quarks and leptons could contain approximate
global symmetries, other than parity,  which would prevent four fermion contact terms from
contributing to atomic parity violation. It would be
interesting to reanalyze the effects of contact terms on physics at
the various colliders,
assuming the 
relations of eq.~\ref{cancel} are satisfied. With SU(2) gauge invariance
\be{\eta^u_{iL}\over (\Lambda^u_{iL})^2
}={\eta^d_{iL}\over (\Lambda^d_{iL})^2}\ee
(neglecting quark CKM mixing)
and so only two independent contact terms need be considered.

 \acknowledgements

This work was supported in part by the DOE under grant \#DE-FG03-96ER40956.

\end{document}